\documentclass[11pt]{article}

\newcommand{\Tr}{\mathrm{Tr}}

\begin{document}

\title {\LARGE A Note on the Generalization of the GEMS Approach}

\author{Hong-Zhi Chen$^{1}$\thanks{Email: hzhchen@pku.edu.cn}, Yu Tian$^{2}$\thanks{Email:
ytian@itp.ac.cn}\\
\\
$^{1}$ School of Physics, Peking University,\\
Beijing 100871, P. R. China\\
$^{2}$ Institute of Theoretical Physics, Chinese Academy of Sciences,\\
P.O. Box 2735, Beijing 100080, P. R. China}

\date{}

\maketitle

\begin{abstract}
This paper is a supplement of our earlier work \cite{Chen:2004qw}.
We map the vector potential of charged black holes into GEMS and
find that its effect on the thermal spectrum is the same as that
on the black hole side, i.e., it will induce a chemical potential
in the thermal spectrum which is the same as that in the charged
black holes. We also argue that the generalization of GEMS
approach to non-stationary motions is not possible.
\end{abstract}

\newpage

\section{Introduction}

It is well known that horizon, radiation and entropy are closely
related objects. The Hawking effect \cite{Hawking:1974sw} is
related to the event horizon intrinsic to the spacetime geometry,
while the Unruh effect \cite{Unruh:1976db} is related to the
horizon associated with a specific observer. However, these two
effects are also closely related. Using a higher dimensional
global embedding Minkowski spacetime (GEMS) of a curved spacetime,
Deser and Levin \cite{Deser:1997ri,Deser:1998bb,Deser:1998xb} find
that the temperature detected by a static detector in a curved
spacetime is equal to the temperature detected by the
corresponding detector (a Rindler detector) in the GEMS. In a
preceding paper \cite{Chen:2004qw}, we generalize their work to
detectors in general stationary motions in the spherically
symmetric spacetimes and match the whole thermal spectra including
the chemical potential in the case of uncharged black holes.
However, we note in \cite{Chen:2004qw} that the chemical potential
can not be properly matched for charged black holes, since the
vector potential does not map into the GEMS. In this paper, we
will address this problem and give a natural mapping of the vector
potential into the GEMS and show that the resulting thermal
spectrum is the same as that of the charged black hole. We will
also make some comments on the generalization to non-stationary
motions and argue that the generalization to that cases is not
possible.

\section{Mapping the vector potential into the GEMS}
We consider the simplest case of a RN black hole.\footnote{We only
consider the non-extremal case.} The metric is
\begin{equation}
ds^2=F(r,M,Q)dt^2-F^{-1}(r,M,Q)dr^2-r^2 (d\theta^2+\sin^2\theta
d\phi^2),
\end{equation}
where
\begin{equation}
F(r,M,Q)=1-\frac{2M}{r}+\frac{Q^2}{r^2}.
\end{equation}
And the 1-form vector potential in a specific gauge which is
natural for an outside observer is
\begin{equation}\label{epotential}
A=-\frac{Q}{r}dt.
\end{equation}
It has two horizons at $r_{\pm}=M\pm\sqrt{M^2-Q^2}$. The usual
GEMS approach only embeds the metric into the higher dimensional
space. However, this is not complete for the charged black hole.
As discussed in \cite{Chen:2004qw}, the static detector outside
the black hole will observe a thermal spectrum with chemical
potential associated with the vector potential, while when viewed
in the GEMS it is a standard Rindler detector and will detect a
thermal spectrum without chemical potential. So we must take into
account the vector potential and find a method to encode its
information in the higher dimensional one. The most natural choice
is to push-forward it into the GEMS. We will see that this is a
correct choice, since it induces the desired chemical potential in
the thermal spectrum detected by the Rindler detector in the
higher dimensional space.

The RN black hole can be embedded in a flat $D=7$ space
\cite{Deser:1998xb} with metric
\begin{equation}\label{emmetric}
ds^2=(dz^0)^2-\sum\limits_{i=1}^{5}(dz^i)^2+(dz^6)^2,
\end{equation}
as follows:
\begin{eqnarray}
z^0 &=& \kappa^{-1}\sqrt{F(r,M,Q)}\sinh(\kappa t), \nonumber \\
z^1 &=& \kappa^{-1}\sqrt{F(r,M,Q)}\cosh(\kappa t), \nonumber \\
z^2 &=& \int\Bigg(\frac{r^2(r_{+}+r_{-})+r^2_{+}(r+r_{+})}{r^2(r-r_{-})}\Bigg)^{1/2}dr, \nonumber \\
z^3 &=& r\sin\theta \cos\phi, \label{rnembed} \\
z^4 &=& r\sin\theta \sin\phi, \nonumber \\
z^5 &=& r\cos\theta, \nonumber \\
z^6 &=&
\int\Bigg(\frac{4r^5_{+}r_{-}}{r^4(r_{+}-r_{-})^2}\Bigg)^{1/2}dr,
\nonumber
\end{eqnarray}
where
\begin{equation}
\kappa=\kappa(r_{+})=\frac{r_{+}-r_{-}}{2r^2_{+}}
\end{equation}
is the surface gravity of the outer horizon. For the cases we are
interested in ($r,\theta=\mathrm{const.}$), the coordinates
$z^2,z^5,z^6$ are constants, and the motions in the higher
dimensional space are effectively four dimensional ones in the
usual Minkowski spacetime (with coordinates $z^0, z^1, z^3, z^4$).
In the following, we only consider this subspace. So in the GEMS,
there are two coordinate systems. One uses $z^0$ as time
coordinate and is the usual Minkowski coordinate system which can
be extended to cover the entire spacetime (when including the
other dimensions) and corresponds to the maximal Kruskal extension
of RN black hole; the other one uses $t$ as time coordinate and is
a Rindler coordinate system which has a horizon at $r=r_+$ and can
only cover the region $r>r_+$. It should be emphasized that the
above embedding can be extended to cover $r<r_{+}$
\cite{Hong:2003xz} so that the observer restricted to $r>r_{+}$
will lost information and hence detect radiation. For a detector
outside the RN black hole moving according to constant $r$,
$\theta$ and $\phi$ with $r>r_+$, the corresponding detector in
the GEMS can be read from eq.(\ref{rnembed}), and it is just a
Rindler detector with a coordinate acceleration $a=\kappa$. The
vector potential (\ref{epotential}), when push-forwarded to the
GEMS, becomes
\begin{eqnarray}
\tilde{A}&=&-\frac{Q}{r}dt \label{rindler} \\
&=&-\frac{1}{\kappa}\frac{z^1}{(z^1)^2-(z^0)^2}\frac{Q}{r}dz^0
+\frac{1}{\kappa}\frac{z^0}{(z^1)^2-(z^0)^2}\frac{Q}{r}dz^1.
\label{minkowski}
\end{eqnarray}
Here eq.(\ref{rindler}) is the form in the Rindler coordinates,
while eq.(\ref{minkowski}) is the form in the Minkowski
coordinates. Following the spirit of GEMS approach, we would like
to calculate the thermal spectrum detected by the Rindler detector
of the Rindler coordinates in the Minkowski vacuum. Because of the
presence of the vector potential, it is simplest to use thermal
Green function method
\cite{Gibbons:1976es,Gibbons:1976pt,Unruh:1983ac}.

We consider a charged scalar field $\phi(x)$ with charge $q$. The
thermal Green function of a grand canonical ensemble with inverse
temperature $\beta$ is defined as
\begin{equation}
G_\beta(x,x')=\langle T\phi(x)\phi(x')^\dag\rangle_\beta,
\end{equation}
where $T$ is the time-ordering operator and the ensemble
expectation value is given by
\begin{equation}
\langle\Phi\rangle_\beta=\frac{\Tr[e^{-\beta(H-\mu
N)}\Phi]}{\Tr[e^{-\beta(H-\mu N)}]},
\end{equation}
where $\mu$ is the chemical potential, $H$ is the Hamiltonian, $N$
is the number operator, and ${\rm Tr}$ means taking trace over a
group of complete bases. Note, in the above definition, that the
time evolution of the field is governed by the usual Heisenberg
equation, i.e.,
\begin{equation}
\phi(t)={e}^{{\rm i}Ht}\phi(0){e}^{-{\rm i}Ht}.
\end{equation}
It is well known that the thermal Green function defined above has
the following quasi-periodicity in imaginary time
\cite{Gibbons:1976pt}:
\begin{equation}
G_\beta(t-t'+{\rm i}\beta)={e}^{\beta \mu}G_\beta(t-t').
\end{equation}
We will now consider the zero temperature Green function of
Minkowski space and show that it just corresponds to the thermal
Green function in Rindler coordinates.

First, we note that the the vector potential (\ref{minkowski}) in
the Minkowski coordinates is singular on the horizon of the
Rindler space $r=r_+$ which corresponds to $(z^1)^2-(z^0)^2=0$. In
order to give a well-behaved vector potential on the horizon, we
perform the following gauge transformation:
\begin{equation}
{\tilde A}'={\tilde A}+d\chi,
\end{equation}
\begin{equation}
\phi'(x)=e^{{\rm i}q\chi(x)}\phi(x),
\end{equation}
with
\begin{equation}
\chi(x)=\frac{Q}{r_+}t.
\end{equation}
Now in this ${\tilde A}'$ gauge the zero temperature Green
function of Minkowski spacetime is suitably analytic in the
Minkowski coordinates \cite{Unruh:1983ac,Hartle:1976tp}, and when
expressed in terms of the Rindler coordinates, it is easy to see
from eq.(\ref{rnembed}) that
\begin{equation}
G^{{\tilde A}'}(t-t'+{\rm i}\beta)=G^{{\tilde A}'}(t-t'),
\end{equation}
with
\begin{equation}\label{temp}
\beta=\frac{2\pi}{\kappa}.
\end{equation}
However, the usual Rindler observers (mapped from the usual
observers outside the RN black hole in the usual gauge) will
describe their observations using the Rindler coordinates and the
${\tilde A}$ gauge, so we should change back to the ${\tilde A}$
gauge and get
\begin{eqnarray}
G^{\tilde A}(t-t') &=& {e}^{-{\rm i}q\chi(x)}{e}^{{\rm
i}q\chi(x')}G^{{\tilde A}'}(t-t') \nonumber\\
&=& {\exp}\Big[-\frac{{\rm i}q Q}{r_+}(t-t')\Big]G^{{\tilde
A}'}(t-t').
\end{eqnarray}
Now it is easy to see that
\begin{equation}
G^{\tilde A}(t-t'+{\rm i}\beta)={e}^{\beta \mu}G^{\tilde A}(t-t'),
\end{equation}
with chemical potential
\begin{equation}
\mu=\frac{q Q}{r_+},
\end{equation}
and $\beta$ still given by eq.(\ref{temp}). This is just the same
thermal spectrum detected by a static detector outside the RN
black hole.

Furthermore, if the detector outside the black hole follows the
trajectory
\begin{equation}\label{traj}
r=r_0,\quad \theta=\theta_0,\quad \phi=\Omega t,
\end{equation}
where $r_0$, $\theta_0$ and $\Omega$ are constants, then as
discussed in \cite{Chen:2004qw}, the detector will detect an
average number of particles with energy $\omega$ as follows
\begin{equation}\label{N}
N_\omega=\frac{1}{{e}^{\beta (\omega-\mu')}-1},
\end{equation}
with chemical potential
\begin{equation}\label{potential}
\mu'=\frac{q Q}{r_+}-m\Omega,
\end{equation}
and $\beta$ given by eq.(\ref {temp}), where $m$ is the orbital
angular quantum number of the particles.

To see the corresponding spectrum in the GEMS, we simply use the
same arguments as above. We see from eq.(\ref{rnembed}) that the
trajectory (\ref{traj}) is mapped into the GEMS as a
$4$-dimensional Rindler motion superposed with a circular motion
perpendicular to the acceleration of the Rindler motion. To go to
the rest frame of the detector, we just perform the coordinate
transformation
\begin{equation}\label{cotr}
{\tilde \phi}=\phi-\Omega t.
\end{equation}
We now focus on a mode with angular quantum number $m$. Its zero
temperature Green function in the Minkowski spacetime is suitably
analytic in the original coordinates and in the ${\tilde A}'$
gauge, and takes the following form
\begin{equation}
G^{{\tilde A}'}_{m}(t-t',\phi-\phi')=F^{{\tilde
A}'}_{m}(t-t'){e}^{{\rm i}m (\phi-\phi')}.
\end{equation}
It will be periodic in $t-t'$, namely we have
\begin{equation}
F^{{\tilde A}'}_{m}(t-t'+{\rm i}\beta)=F^{{\tilde A}'}_{m}(t-t')
\end{equation}
with $\phi-\phi'$ invariant and $\beta$ given by eq.(\ref{temp}).
The embedding observers will describe their observations using
coordinate ${\tilde \phi}$ and the ${\tilde A}$ gauge, so after
change back to ${\tilde A}$ gauge and perform the coordinate
transformation (\ref{cotr}), we get
\begin{equation}
G^{{\tilde A}}_{m}(t-t',{\tilde\phi}-{\tilde\phi}')=F^{{\tilde
A}'}_{m}(t-t'){e}^{{\rm i}m ({\tilde\phi}-{\tilde\phi}')}{e}^{{\rm
i}m\Omega (t-t')}{e}^{-{\rm i}q Q(t-t')/{r_+}}
\end{equation}
Taking ${\tilde \phi}-{\tilde \phi}'$ invariant, it is easy to see
that
\begin{equation}
G^{{\tilde A}}_{m}(t-t'+{\rm i}\beta,{\tilde \phi}-{\tilde
\phi}')={e}^{\beta {\tilde \mu}}G^{{\tilde A}}_{m}(t-t',{\tilde
\phi}-{\tilde \phi}'),
\end{equation}
with $\beta$ again given by eq.(\ref{temp}) and chemical potential
given by
\begin{equation}
{\tilde \mu}=\frac{q Q}{r_+}-m\Omega.
\end{equation}
This is just the same thermal spectrum
eq.(\ref{N},\ref{potential}) detected by a rotating detector
outside the RN black hole.

\section{Comments on the Generalization to Non-Stationary motions}

The above discussions together with results of \cite{Chen:2004qw}
indicate that the GEMS approach can be generalized to general
stationary motions in curved spacetimes. It is interesting to ask
whether this approach can be generalized to even more general
motions, i.e., non-stationary motions. One reason of this
generalization is the hope of using the techniques available in
the quantum field theory on flat Minkowski spacetime to study the
more complicated issues in curved spacetimes. In fact, this is one
of our motivations to study this approach. However, we will argue
that the generalization to non-stationary motions is not possible.

First, we note that in some cases, the global embedding space of a
curved spacetime will have more than one time dimension. Take RN
black hole for example. Its embedding space (\ref{emmetric}) has
an extra time dimension $z^6$ which only depends on $r$. For a
non-stationary motion with varying $r$, we have to deal with a
quantum field theory on a spacetime with two times which we even
do not know how to define it. On the other hand, although the
non-stationary motion in the curved spacetime is complicated, we
can still deal with it using technique of particle detector. So in
this case, the generalization is not possible.

For the Schwarzschild spacetime whose embedding space is an
ordinary $D=6$ Minkowski spacetime with only one time dimension,
the situation is not better either. For example, suppose we want
to consider a freely falling detector outside a Schwarzschild
black hole. Since a static detector in the curved space maps to a
static detector in the Rindler space, and a rotating detector maps
to a rotating detector in the Rindler space, naively, we will
expect a freely falling detector maps to a freely falling detector
in the Rindler space. Furthermore, a freely falling detector in
the Rindler space is just an inertial detector in the Minkowski
space. For such a detector, it is well known that it will detect
nothing in the Minkowski vacuum. So, following the spirit of the
GEMS approach, it seems that we may conclude that a freely falling
detector in the Hartle-Hawking vacuum in a Schwarzschild spacetime
will detect nothing. Indeed, it is known that a freely falling
detector sees essentially no particles near the horizon
\cite{unruh77}. It seems that we have matched the two sides.
Unfortunately, the naive expectation is not correct. Generally
speaking, the embedding does not preserve geodesics. The simplest
example is that a geodesic (a large circle) on a 2-sphere $S^2$ is
not a geodesic (a straight line) in its embedding Euclidean space
$E^3$. For the case we are interested in, this can be seen from
the global embedding of the Schwarzschild spacetime. It can be
embedded in a $D=6$ Minkowski spacetime with metric
\begin{equation}\label{6dm}
ds^2=(dz^0)^2-(dz^1)^2-(dz^2)^2-(dz^3)^2-(dz^4)^2-(dz^5)^2,
\end{equation}
as follows \cite{fronsdal}:
\begin{eqnarray}
z^0 &=& 4M\sqrt{1-\frac{2M}{r}}\sinh\Big(\frac{t}{4M}\Big), \nonumber \\
z^1 &=& 4M\sqrt{1-\frac{2M}{r}}\cosh\Big(\frac{t}{4M}\Big), \nonumber \\
z^2 &=& \int dr \sqrt{\frac{2Mr^2+4M^2r+8M^3}{r^3}}, \nonumber \\
z^3 &=& r\sin\theta \sin\phi, \nonumber \\
z^4 &=& r\sin\theta \cos\phi, \nonumber \\
z^5 &=& r\cos\theta. \label{schembed}
\end{eqnarray}
On the other hand, for a freely falling observer outside the
Schwarzschild black hole, its rest frame ($\tau, R$) is related to
the Schwarzschild coordinates $(t, r)$ through the Lemaitre
transformation \cite{lemaitre} as follows:
\begin{eqnarray}\label{lemaitre}
r&=&(2M)^{1/3}\Big[\frac{3}{2}(R-\tau)\Big]^{2/3}, \nonumber\\
t&=&\tau-2(2M)^{2/3}\Big[\frac{3}{2}(R-\tau)\Big]^{1/3}\nonumber\\
&&-2M\ln \Bigg
|\frac{\big[\frac{3}{2}(R-\tau)\big]^{1/3}-(2M)^{1/3}}
{\big[\frac{3}{2}(R-\tau)\big]^{1/3}+(2M)^{1/3}}\Bigg|,
\end{eqnarray}
with $\theta$, $\phi$ invariant. A freely falling observer
corresponds to $(R,\theta,\phi)=\mathrm{const}$. From
eqs.(\ref{schembed},\ref{lemaitre}), it can be shown by
straightforward calculations that the freely falling observer in
the Schwarzschild spacetime does not map to the inertial observer
in the embedding Minkowski spacetime. In fact, we can also
anticipate this by noting that the geodesic motion in the
Minkowski space is stationary rather than non-stationary. On the
contrary, the embedding motion is complicated and the
corresponding detector will detect particles. Thus, we see that
even for this relatively simple non-stationary motion, the results
of the two sides cannot match each other. So we conclude that the
generalization of the GEMS approach to non-stationary motions is
not possible.

\section{Conclusion}

We see that after properly mapping the vector potential of the RN
black hole into the GEMS, we can match the whole thermal spectrum
completely, including the chemical potential, for both static and
rotating detectors. The generalization to other spherically
symmetric charged black holes, such as RN-dS and RN-AdS black
holes, and the generalization to other dimensions are
straightforward. Thus we have confirmed that the proposal of Deser
and Levin in \cite{Deser:1997ri,Deser:1998bb,Deser:1998xb} is
valid in these stationary motions. We also argue that the
generalization to non-stationary motions is not possible. This
means that the GEMS approach can be only applied to  stationary
motions in curved spacetime.

\section*{Acknowledgements}

We would like to thank Prof. Yi~Ling, Doctors Z.-M.~Hu, H.-T.~Lu,
H.-L.~Wang for helpful discussions.

\end{document}